\documentclass[12pt]{article}

\usepackage{epsfig}
\usepackage{amsfonts}
\usepackage{amsopn}
\usepackage{amsmath}
\usepackage{verbatim,amsthm}

\begin{document}

\title
{
\vskip-50 pt
\begin{flushright}
\normalsize\rm NORDITA-2017-070
\end{flushright}
\vskip 20 pt
Inflation versus collapse in brane matter
}
\author
{
 A. A. Zheltukhin $^{a,b}$\thanks{e-mail: aaz@fysik.su.se
 } 
 \\\\
$^a$ Kharkov Institute of Physics and Technology, 
Kharkov, 61108, Ukraine \\  
$^b$ Nordita, KTH Royal Institute of Technology and Stockholm University\\
SE 106 91 Stockholm, Sweden
}
\maketitle

\begin{abstract}

 Mapping of fundamental branes to  
  their worldsheet (ws) multiplets originating  from 
  spontaneous breaking of the Poincare symmetry is studied. 
The interaction Lagrangian for fields of the 
 Nambu-Goldstone multiplet is shown to encode $R^2$ gravity on the 
  ws. The power law $k_{p}\sim T_{p}^{\frac{3-p}{2(p+1)}}$ 
 for the SO(D-p-1) gauge coupling $k_{p}$ as the function of the p-brane 
 tension $T_{p}$ is assumed.  It points to the presence of 
  asymptotic freedom and confinement phases in brane matter. 
Their connection  with collapse and inflation of the branes is discussed.

\end{abstract}

\section{Introduction}

The geometric approach  \cite{RL}, \cite{BN} reduces search for
 fundamental Nambu-Goto  strings and Dirac branes to studying 
the Plateu problem for hyper-surfaces 
in pseudo-Riemannian spaces \cite{Dirb}, \cite{Eisn}. 
The problem is investigated by the Cartan method 
of moving frames used in the theory of Lie groups
 and symmetric spaces \cite{Car}, \cite{JS}.
 This method was applied for description 
 of the Nambu-Goldstone (NG) bosons in systems with a spontaneously 
 broken internal symmetry nonlinearly realized by a group $G$
 \nocite{Wei, Schw, Vol1, CWZ, CCWZ} [7-11]. 
   Interpretation of the NG bosons  as the constrained 
multiplets of the gauge symmetry $H$, arisen from localizing 
the global vacuum symmetry $H\in G$, was proposed in Ref. \cite{FST}.
 These ideas were used for investigation of $p$-dimensional  
{\it minimal} hyper-worldsheet (ws) $\Sigma_{p+1}^{min}$ swept by 
strings $(p=1)$ \cite{Zgau2} or $p$-branes 
\cite{Z_rmp}, \cite{Zgau_p} embedded into the $D$-dimensional Minkowski space. 
Thereat, the space-time Poincare group $ISO(1, D-1)$ 
was treated as the brane symmetry spontaneously broken to 
 $ISO(1,p)\times SO(D-p-1)$ initiated by the  
  embedding of  $\Sigma_{p+1}^{min}$ 
  \cite{GKW}, \cite{GKP}, \cite{BZ_hedr}.  
 The gauge and diff-invariant field action considered  in \cite{Z_rmp}
  includes  the $SO(D-p-1)$ gauge 
 multiplet $B_{\mu}^{ab}$ interacting with 
 the  NG tensor multiplet $l_{\mu\nu}^{a}$ 
  and the ws metric $g_{\mu\nu}$ on $\Sigma_{p+1}^{min}$.
  
 Here we find that the action \cite{Z_rmp}
  gives an implicit ws realization 
 of $(p+1)$-dimensional gravity described by quadratic curvature 
 terms similar to those used in Refs. [19-23] \nocite{Stel, Strb, Frad, Zwi, LPPS}. 
 These terms are encoded by the quartic  potential 
for the field $l_{\mu\nu}^{a}$ identified with the second fundamental
 form of $\Sigma_{p+1}^{min}$. For the codimension 1, this
 reveals an alternative formulation of the $R^{2}$ gravity \cite{Strb}.
 The latter also contains a cosmological term represented by an  
integation constant $\Lambda$. 

On the other hand, we understand the field action  
to map the Dirac brane action into the one which 
describes the Yang-Mills (YM) and tensor ws multiplets. 
The map unambiguity demands addition of the Ricci-Codazzi 
equations as the initial  data 
for the Euler-Lagrange variational equations 
(see proof in Ref. \cite{Z_rmp}). 
If $\Lambda=0$ the action has only one coupling 
constant $k_{p}$ depending of the brane 
tension $T_{p}$ by means of the power law 
$k_{p}\sim T_{p}^{\frac{3-p}{2(p+1)}}$. 
This shows that $k_{p}$ treated as the function of $T_{p}$
has three different regimes of behavior corresponding
 to the cases $p<3, p=3$ and $p>3$. Taking into account
  that the value of tension defines the energy scale,
 one can estimate how  $k_{p}$ changes with energy.
 Then we explain the regimes as the ones corresponding to 
the asymptotic freedom or confinement.
Using the w-s curvature dependence on $k_{p}$
 we find that the field regimes describe the inflation or collapse 
 phases of the branes \footnote{Some exact solutions describing collapsing p-branes 
 were found in Refs. \cite{TrZ} and \cite{Zcol}.}. 
 For  $p=3$ the coupling $k_{3}$ 
 becomes dimensionless and independent of 
  $T_{3}$.  Then the field action  
  becomes scale invariant in correspondence with the results
  obtained in [26-29] \nocite{Dir, DFF, Smol, Zee} 
  (see also Ref. \cite{Perv}).

\section{Branes and quadratic curvature gravity}

In string theory, a (p+1)-dimensional hyper-ws 
embedded into D-dimensional
 Minkowski space $\mathbf{R}^{1,D-1}$ 
is described by its worldvector $\mathbf{x}(\xi^{\rho})$ 
depending on the internal coordinates $\xi^{\mu}=(\tau,\sigma^r), \, r=1,2,..,p$.
In the Cartan approach \cite{Car} 
to differential geometry, a (p+1)-dimensional hyper-surface $\Sigma_{p+1}$ 
is described by the Maurer-Cartan or Gauss-Ricci-Codazzi (G-R-C) 
equations \cite{Eisn} that can be 
rewritten as the covariant field constraints  studied in Ref.\cite{Z_rmp}.

The Gauss constraints  for the Riemann tensor $R_{\mu\nu\gamma\lambda}(\xi)$ 
of $\Sigma_{p+1}$
 \begin{eqnarray}\label{cRl}
R_{\mu\nu}{}^{\gamma}{}_{\lambda}=l_{[\mu}{}^{\gamma a} l_{\nu]\lambda a}
\end{eqnarray} 
express this tensor through the second fundamental form $l_{\mu\nu}^{a}(\xi)$. 
The indexes $a,b=p+1,p+2,..., D-1$ enumerate   
the orts $\mathbf{n}_{a}(\xi^{\rho})$ of an orthonormal moving frame
 attached to $\Sigma_{p+1}$ which are orthogonal to it.  
The symmetric tensor $l_{\mu\nu}^{a} $ is treated as a constrained multiplet 
of the local group $SO(D-p-1)$ 
 \footnote{All notations and definitions coincide with the ones used in Ref. \cite{Z_rmp}.}. 

The Ricci equations are equivalent to the constraints 
\begin{eqnarray}\label{cH2}
H_{\mu\nu }{}^{ab}= l_{[\mu}{}^{\gamma a} l_{\nu]\gamma}{}^{b}, \ \ \ \   
H_{\mu\nu a}{}^{b}:=(\partial_{[\mu}B_{\nu]} + [B_{\mu}, B_{\nu}])_{a}{}^{b}
\end{eqnarray}
for the field strength $H_{\mu\nu a}{}^{b}$  
of the  $SO(D-p-1)$ gauge field $B_{\mu}^{ab}(\xi)=-B_{\mu}^{ba}(\xi)$ 
 in the fundamental representation.
 The brackets  $\{\mu\nu\},\, [\mu\nu]$ imply the $\mu,\nu$ 
symmetrization and antisymmetrization, respectively. 
Eqs.  (\ref{cH2}) appear  only for the 
codimension $D-(p+1)\geq2$,  otherwise $B_{\mu a}{}^{b}\equiv 0$.

The Codazzi equations are equivalent to the field constraints 
\begin{eqnarray}\label{ccd'}
\nabla_{[\mu}^{\perp}l_{\nu]\rho}^{a}=0,
\end{eqnarray}
where $\nabla_{\mu}^{\perp}l_{\nu\rho}{}^{a}$ is the metric and YM  
 covariant derivative 
\begin{eqnarray}\label{cdl}
\nabla_{\mu}^{\perp}l_{\nu\rho}{}^{a}:= \partial_{\mu}l_{\nu\rho}{}^{a}
- \Gamma_{\mu\nu}^{\lambda} l_{\lambda\rho}{}^{a} 
-\Gamma_{\mu\rho}^{\lambda} l_{\nu\lambda}{}^{a} + B_{\mu}^{ab}l_{\nu\rho b}.
 \end{eqnarray} 
The corresponding Bianchi identities have the form
  \begin{eqnarray}\label{BIl}
[\nabla^{\perp}_{\gamma} , \, \nabla^{\perp}_{\nu}] l^{\mu\rho a}
=R_{\gamma\nu}{}^{\mu}{}_{\lambda} l^{\lambda\rho a}  
+ R_{\gamma\nu}{}^{\rho}{}_{\lambda} l^{\mu\lambda a} 
+H_{\gamma\nu}{}^{a}{}_{b} l^{\mu\rho b}.  
\end{eqnarray}
The GRC eqs. (\ref{cRl}-\ref{ccd'}) complemented by 
the minimality conditions for  $\Sigma_{p+1}$
\begin{eqnarray}\label{minco}
Spl^{a}:= g^{\mu\nu}l^{a}_{\nu\mu}=0
\end{eqnarray}
 form a complete set of the equations of motion (EOM) for the Dirac p-brane embedded 
into  $\mathbf{R}^{1,D-1}$. 

The $SO(D-p-1)$ and diff-invariant action of a p-brane 
sweeping a {\it minimal} hyper-ws $\Sigma^{min}_{p+1}$ and  
consistent with (\ref{cRl}-\ref{ccd'}) has the form \cite{Z_rmp}
 \begin{eqnarray} 
S_{Dir}= \frac{1}{k_{p}^{2}}\int d^{p+1}\xi\sqrt{|g|}\, 
\{- \frac{1}{4}Sp(H_{\mu\nu}H^{\nu\mu}) \nonumber \\
+ \, \frac{1}{2}\nabla_{\mu}^{\perp}l_{\nu\rho a}\nabla^{\perp \{\mu}l^{\nu\}\rho a}
-\nabla_{\mu}^{\perp}l^{\mu}_{\rho a}\nabla_{\nu}^{\perp}l^{\nu\rho a} + V_{Dir}(l, g)\}, 
\label{actnl}
\end{eqnarray}
 where $Sp(H_{\mu\nu}H^{\nu\mu}):=H_{\mu\nu}^{ab}H^{\nu\mu}_{ba}$ 
is the trace with respect to the indexes $a, b$.  
 Here we use  $\hbar=c=1$ and the field dimensions  
 $[l_{\mu\nu}^{a}]=[B_{\mu}^{ab}]= [\nabla^{\perp}_{\nu}]
 =[L^{-1}], \,  [\xi^{\mu}]=[L], \, [g_{\mu\nu}]=1$ resulting in
  the dimension of the coupling $k_{p}$ 
 \begin{equation}\label{kdim}
[k_{p}]=[L^{\frac{d_{p}-4}{2}}], \ \ \ \ [{\rm c_{p}}]=[L^{-4}], \ \ \ \ d_{p}:= p+1.
\end{equation}  
 The diff-invariant potential $V_{Dir}(l, g))$ encodes self interaction 
 of the NG multiplet $l_{\mu\nu}^{a}$ 
 in the gravitational {\it background} $g_{\mu\nu}(\xi^{\rho})$
\begin{eqnarray}\label{solVl} 
V_{Dir}=- \frac{1}{2} Sp(l_{a}l_{b}) Sp(l^{a}l^{b})
+ Sp(l_{a}l_{b}l^{a}l^{b}) - Sp(l_{a}l^{a}l_{b}l^{b})+ {\rm c_{p}}, 
\end{eqnarray}
where ${\rm c_{p}}$ is an integration constant. 
The Euler-Lagrange PDEs following from  (\ref{actnl}) have
 a unique solution that describes the fundamental branes 
provided that Eqs. (\ref{cH2}-\ref{ccd'}) are chosen as  
 the {\it Cauchy initial data}.
The latter  considered as the functions of the proper 
time $\tau$ turned out to be invariants of the evolution 
equations for of $l^{a}_{\mu\nu}, \, B_{\mu}^{ab}$ following from 
$S_{Dir}$ (see proof in Ref\cite{Z_rmp}). 

The metric dynamics is described by Eqs. (\ref{cRl})
treated as the evolution PDEs for $g_{\mu\nu}$.
 These equations are consistent with the used variational 
 principle since they have selected $V_{Dir}$ 
 which can be rewritten in the form 
 \begin{eqnarray}\label{R2trm}
V_{Dir}=-\frac{1}{4}R_{\mu\nu\gamma\lambda}R^{\mu\nu\gamma\lambda}
-\frac{1}{2}R_{\mu\nu}R^{\mu\nu}+ \frac{1}{4}H_{\mu\nu ab }H^{\mu\nu ab} 
+{\rm c_{p}}.
\end{eqnarray} 
 Relation (\ref{R2trm}) was derived using Eqs.(\ref{cRl}-\ref{cH2}) and (\ref{minco})
  resulting in the relations 
\begin{eqnarray}
\frac{1}{2}R_{\mu\nu\gamma\lambda}R^{\mu\nu\gamma\lambda}
=Sp(l_{a}l_{b})Sp(l^{a}l^{b}) - Sp(l_{a}l_{b}l^{a}l^{b}),  \label{rim2}
\\
\frac{1}{2}H_{\mu\nu}^{ab}H^{\mu\nu}_{ab}
=Sp(l_{a}l_{b}l^{a}l^{b}) - Sp(l_{a}l^{a}l_{b}l^{b}), \, \, \, \, \, \, Spl^{a}=0,
\label{max}
\end{eqnarray}
which were combined with the quadratic expressions for 
 the Ricci tensor $R_{\mu\nu}$ and the scalar curvature $R$ of 
the  $\it minimal$ hyper-ws $\Sigma_{p+1}^{min}$ 
\begin{eqnarray}\label{HiEi}
R_{\mu\nu}=- (l^{a}l_{a})_{\mu\nu},  \, \, \, \, \, R=-Sp(l^{a}l_{a}).  
\end{eqnarray}
The brane potential (\ref{R2trm}) contains the curvature squared terms 
similar to those in the action of string theory \cite{Zwi}. 
In the codimension 1, i.e. when  $D=p+2$, the field $B_{\mu}^{ab}\equiv 0$ 
since $a=b=p+1$ 
and  (\ref{actnl}) reduces to the action 
\begin{eqnarray}\label{actncol}
S_{D=p+2} =\frac{1}{k_{p}^{2}}\int d^{p+1}\xi\sqrt{|g|}
(\frac{1}{2}\nabla_{\mu}l_{\nu\rho  \perp}\nabla^{\{\mu}l^{\nu\}\rho \perp}  \\ 
-\nabla_{\mu}l^{\mu}_{\rho \perp}\nabla_{\nu}l^{\nu\rho\perp}
 - \frac{1}{2} [Sp(l_{\perp}l^{\perp})]^2 + {\rm c_{p}}), 
\nonumber 
 \end{eqnarray}
 where $p+1$  is  denoted as $\perp$ and the pure metric covariant
  derivative as $\nabla_{\mu}$ 
\begin{eqnarray}\label{cdlco}
\nabla_{\mu}l_{\nu\rho}^{\perp}:= \partial_{\mu}l_{\nu\rho}^{\perp}
- \Gamma_{\mu\nu}^{\lambda} l_{\lambda\rho}^{\perp} 
-\Gamma_{\mu\rho}^{\lambda} l_{\nu\lambda}^{\perp},  \ \ \ \ \
 l_{\lambda\rho}^{\perp}=- l_{\lambda\rho\perp}\equiv -l_{\lambda\rho}.
 \end{eqnarray}  
Then, the quadratic terms in (\ref{actncol}) reduce to 
one half of the squared curvature 
\begin{eqnarray}\label{cur2}
\frac{1}{2}[Sp(l_{\perp}l^{\perp})]^{2}= \frac{1}{2}R^{2}  
\end{eqnarray}
in view of the relations (\ref{R2trm}-\ref{HiEi}). 
This result extends the statement \cite{GM}  
that the first leading correction to  ${\it string}$ 
action is quadratic in $R$, to  the Dirac $p$-branes 
with codimensions equal to 1 
\footnote{For the {\it open} strings the term linear in $R$ 
contributes to the boundary conditions for the EOM uncovering
a hidden topological structure of the action extremals \cite{Zbc}.}.

So, the discussed action $S_{Dir}$ encodes the terms 
describing $R^2$ gravity on the string/brane ws in
the potential $V_{Dir}$ of the N-G fields $l_{\mu\nu}^{a}$.

 \section{Phases of brane matter}

The considered approach maps the known 
Dirac brane action \cite{Dirb}
\begin{equation}\label{DNG}
S=T_{p}\int d^{p+1}\xi 
\sqrt{|det(\partial_{\mu}\mathbf{x}\partial_{\nu}\mathbf{x})|}
\end{equation}
 into the field action (\ref{actnl}) 
consistent with Eqs.(\ref{cRl}-\ref{ccd'}) and containing 
 one coupling constant $k_{p}$ if ${\rm c_{p}}=0$. 
 Then $k_{p}$  must be a function of the tension $T_{p}$. 
 
Choosing the dimension $[\mathbf{x}]=[\xi^{\mu}]=[L]$, 
we find that
\begin{equation}\label{Tkdim}
[T_{p}]=[L^{-d_{p}}], \, \, \, \, [k_{p}]
=[L^{\frac{d_{p}-4}{2}}]=[T_{p}]^{\frac{4-d_{p}}{2d_{p}}} 
\end{equation} 
using (\ref{kdim}). Transition to the fields with the canonical
 dimension $[L^{\frac{2-d_{p}}{2}}]$
\begin{eqnarray}\label{regrad}
{\tilde l}_{\mu\nu}^{a}:=k_{p}^{-1}l_{\mu\nu}^{a},  \, \, \, \, \, \, \, \, \, 
{\tilde B}_{\mu}^{a b}:=k_{p}^{-1}B_{\mu}^{a b} 
 \end{eqnarray} 
 shows that $k_{p}$ coincides with the gauge coupling 
 for the group $SO(D-p-1)$:
\begin{eqnarray}
\nabla_{\mu}^{\perp}{\tilde l}_{\mu\nu}^{a}= k_{p}^{-1}\nabla_{\mu}^{\perp} l_{\nu\rho}{}^{a}  
\equiv\partial_{\mu}{\tilde l}_{\nu\rho}{}^{a}
- \Gamma_{\mu\nu}^{\lambda}{\tilde l}_{\lambda\rho}{}^{a} 
-\Gamma_{\mu\rho}^{\lambda}{\tilde l}_{\nu\lambda}{}^{a} 
+ k_{p}{\tilde B}_{\mu}^{ab}{\tilde l}_{\nu\rho b}, 
\nonumber \\ 
{\tilde H}_{\mu\nu }{}^{ab}=k_{p}^{-1}H_{\mu\nu }{}^{ab}
\equiv (\partial_{[\mu}{\tilde B}_{\nu]} + k_{p} {\tilde B}_{[\mu}{\tilde B}_{\nu]})^{ab}
\, \, \, \, \, \, \, \, \, \, \, \, \, \, \, \, \, \,
\label{redef}
\end{eqnarray}
for $D\geq p+3$ when  ${\tilde B}_{\nu}^{ab}\neq 0$.
 
 In this case, $k_{p}$ squared  is equal to the 
 interaction coupling $\lambda_{p}$ in $V_{Dir}({\tilde l})$  
 \begin{eqnarray}\label{conect}
\lambda_{p} = k_{p}^{2}
\end{eqnarray}
In terms of the canonical fields, the action (\ref{actnl}) takes the form 
\begin{eqnarray} 
S_{Dir}= \int d^{p+1}\xi\sqrt{|g|}\, \{- \frac{1}{4}Sp({\tilde H}_{\mu\nu}{\tilde H}^{\nu\mu}) \nonumber \\
+ \, \frac{1}{2}\nabla_{\mu}^{\perp}{\tilde l}_{\nu\rho a}\nabla^{\perp \{\mu}{\tilde l}^{\nu\}\rho a}
-  \nabla_{\mu}^{\perp}{\tilde l}^{\mu}_{\rho a}\nabla_{\nu}^{\perp}{\tilde l}^{\nu\rho a} \label{actn} \\ \nonumber
+ \, k_{p}^{2}\ [- \frac{1}{2} Sp({\tilde l}_{a}{\tilde l}_{b}) Sp({\tilde l}^{a}{\tilde l}^{b})
+ Sp({\tilde l}_{a}{\tilde l}_{b}{\tilde l}^{a}{\tilde l}^{b}) - Sp({\tilde l}_{a}{\tilde l}^{a}{\tilde l}_{b}{\tilde l}^{b})] + \Lambda_{p} \},
\end{eqnarray}
 where  $\Lambda_{p}:={\rm c_{p}}/k_{p}^{2}$ is a cosmological 
 constant with the dimension $[\Lambda_{p}]=[L^{-d_{p}}]=[T_{p}]$ 
 in view of (\ref{kdim}). When  $\Lambda_{p}=0$, dimensional analysis implies that 
\begin{eqnarray}\label{Tk}
k_{p}\sim T_{p}^{\frac{4-d_{p}}{2d_{p}}}\equiv T_{p}^{\frac{3-p}{2(p+1)}}, 
\ \ \ \ \    \lambda_{p}\sim T_{p}^{\frac{3-p}{p+1}}. 
\end{eqnarray}
By analogy with condensed matter physics, one can treat 
 $T_{p}=0$ and $\alpha_{p}$ 
\begin{eqnarray}\label{Crit}
 \alpha_{p}:=\frac{3-p}{2(p+1)},  \,  \ \ \ \ \ \ p=1,2,.., D-1
\end{eqnarray}  
as the critical temperature and exponent, respectively,
depending on tne brane dimension $p$.  
The tensionless limit corresponds to super-Planckian energies  
when the Planck mass becomes negligible
 (see e.g. Refs. \nocite{Shil, KL, Liz, GroM, Znul, BZnul} [33-38]).
 
 The power law (\ref{Tk}) shows that the behavior of 
the coupling constants as the functions of brane tensions
is characterized by three different regimes defined by the sign of  
$\alpha_{p}$. 
They correspond to different phases of the multiplet fields  
 describing the brane matter in  (\ref{actn}).  
 So, we observe the {\it decrease} of the constants in
cases when $p<3$  corresponding to strings and membranes 
$$
k_{1}\sim T_{1}^{\frac{1}{2}}, \ \  \lambda_{1}\sim T_{1}; 
  \ \ \ \ \
k_{2} \sim T_{2}^{\frac{1}{6}}, \ \ \lambda_{2}\sim T_{2}^{\frac{1}{3}} 
$$
and their {\it increase} for the phases corresponding 
to $p$-branes with  $p>3$  
 $$
k_{4}\sim T_{4}^{-\frac{1}{10}}, \ \  \lambda_{4}\sim T_{4}^{-\frac{1}{5}};
  \ \  
k_{5} \sim T_{5}^{-\frac{1}{6}}, \ \  \lambda_{5}\sim T_{1}^{-\frac{1}{3}}; 
$$
when $T_{p}$ {\it decreases} together 
with  $k_{\infty} \sim T_{\infty}^{-\frac{1}{2}},
\ \ \lambda_{\infty}\sim T_{\infty}^{-1}$ 
for $D,p\rightarrow \infty$. 

The dependence of the metric and YM curvatures 
of the hyper-ws on $k_{p}$
 follows from Eqs. (\ref{cRl}-\ref{ccd'}) 
rewritten in terms of the canonical fields 
\begin{eqnarray}
R_{\mu\nu}{}^{\gamma}{}_{\lambda}
=k_{p}^{2}{\tilde l}_{[\mu}{}^{\gamma a}{\tilde l}_{\nu]\lambda a}   
\label{R'} \\
  {\tilde H}_{\mu\nu }{}^{ab}= 
  k_{p}{\tilde l}_{[\mu}{}^{\gamma a}{\tilde l}_{\nu]\gamma}{}^{b},
\label{H'} \\ 
\nabla_{[\mu}^{\perp}{\tilde l}_{\nu]\rho a}=0. \ \ \ \ \ \ \ 
\label{Co'}
\end{eqnarray} 
These eqs. show that in super-Planckian region,     
  the string and membrane curvatures go to zero 
  that corresponds to their inflation. 
On the contrary, the curvature of branes with $p>3$ goes to infinity 
that could be treated as their collapse if their rotations do not 
compensate their large tensions.

Thus, at high energies the regimes of 
confinement $(k_{3,4,...}\rightarrow \infty)$
or asymptotic freedom ($k_{1}, k_{2} \rightarrow 0$) 
in the field model (\ref{actn}) correspond to 
brane matter phases describing its collapse or inflation 
depending on the value of $p$.

In the low-energy limit, when  $T_{p}\rightarrow \infty$, 
the curvatures go to infinity for  $p=1,2$ and to zero for 
$p>3$ that could be treated as the  
collapse of non-rotating strings, membranes contrary to  
an inflation of the rotating branes with $p=4,5,.., D-2$.

 Finally, the field regime  when $\alpha_{p}=0$ 
  emerges  at  $p=3$. 
  In this exclusive  case the coupling constant $k_{3}$
 in (\ref{actn}) is dimensionless and independent of $T_{p}$. 
 Since in our analysis  the constant $\Lambda_{3}$ 
 with the dimension $[L^{-4}]$ vanishes  
 the action (\ref{actn}) with $p=3$ becomes 
invaiant under the global Weyl transformatins 
 \begin{eqnarray}\label{scali}
 g'_{\mu\nu}(\xi)=\rho g_{\mu\nu}(\xi), \, \, \, \,
  {\tilde l'}_{\mu\nu}(\xi)=\rho^{1/2} {\tilde l}_{\mu\nu}(\xi), \, \, \, \,
 {\tilde B}_{\mu}^{' ab}(\xi)={\tilde B}_{\mu}^{ab}(\xi).
 \end{eqnarray} 
  So, we come into contact with the well-known conformal invariant theories of gravity
(see Ref. \cite {JoNo} and references therein).

 \section {Summary}

 The correspondence between {\it fundamental} $p$-branes, 
 sweeping {\it minimal} hyper-surfaces 
in D-dimensional Minkowski space, and their $SO(D-p-1)$ gauge 
ws multiplets is studied. 
It is shown that the interaction potential of the 
Nambu-Goldstone multiplet encodes
the ws action of $R^2$ gravity.
 Based on dimensional analysis of the ws action (\ref{actn})
  with zero cosmological constant, we obtain 
 the power law $k_{p}\sim T_{p}^{\frac{3-p}{2(p+1)}}$ 
 for the coupling constant as the function of the $p$-brane 
 tension $T_{p}$. This points to possible existence of brane 
 matter phases interpreted as asymptotic freedom and 
 confinement phases. Their connection with collapse and inflation 
 of $p$-brane hyper-ws is discussed. 
The phase corresponding to $p=3$ describes a 
scale-invariant model of $R^2$ gravity on a three-brane hyper-ws.
These observations could be applied in cosmology. Indeed,   
 action (\ref{actn}) proposes a (p+1)-dimensional field model 
 unifying the $SO(D-p-1)$ YM and NG multiplets with 
 gravity. Then the extended object - a fundamental p-brane - 
 emerges as the solution of the Cauchy-Kowalewskaya problem 
 for the EOM. Consideration of the brane hyper-ws as a p-braneworld
  model for the space-time similarly to \cite {RS} could explain 
  some of the open questions.
 On the other hand, one can consider p-branes as dynamical objects
  filling our universe and forming domains with various dimensions 
 $p=1,2,3,4,.., D-2$. 
 Coalescence and fragmentation of the domains 
 during the evolution could change their dimensions and trigger 
 the above discussed phase transitions in brane matter, respectively.
 
 Treatment of these processes demands consideration 
 of the interaction between the branes.  This interaction can be realized 
 using the Kalb-Ramond action-at-a-distance Lagrangian and the 
 corresponding invariant action for interacting strings \cite{KR} 
 generalized to the case of interacting p-branes \cite{YIIMYY}. 
  The interaction of p-branes is mediated by 
 the gauge field presented by a completely antisymmetric 
 tensor $B_{m_{0} m_{1}...m_{p}}$ of the rank (p+1) \cite{Z_nul} 
 generalizing the Kalb-Ramond gauge boson $B_{mn}$. 
         
The discussed geometric approach can be extended to include 
 fermionic fields using the ideas of supersymmetry and supergravity. 
This way implies extension of the Minkowski space by the Grassmannian 
spinor coordinates $\theta^{\alpha}$ generating  
 fermionic degrees of freedom.  The resulting flat superspace formed by 
 the coordinates $(x^{m},\theta^{\alpha})$ is invariant under the global 
 super-Poincare group. Supersymmetric field theories formulated in the superspace 
 become superfield theories.
  Localization of the super-Poincare group yields the gravity theory including 
  the fermionic Rarita-Schwinger gauge field carrying spin $3/2$. 
General relativity reformulated by Cartan using 
the exterior differential forms as the theory on a group manifold 
was extended to arbitrary superspaces.  
The generalized  Maurer-Cartan structure equations were 
constructed \cite{VS}, \cite{AFR}.  
 The Generalized Action Principle (GAP) together with the rheonomy principle
  formulated in Ref. \cite{AFR} allowed to extend a component formalism 
  of supergravity on the superspace. 
  Application of the GAP complemented by the new principle, 
  called the rheotropy, allowed  to formulate the procedure of minimal embedding
  of super hyper-surfaces of superstrings and super p-branes into the Minkowski
   target superspace (see Ref. \cite{Vol_Gap} and references therein).
 This development of the Cartan's approach to the differential geometry
 of embedded super hyper-surfaces gives tools  for construction of  
 a supersymmetric generalization of the bosonic action $S_{Dir}$ (\ref{actnl}).
 A generalization of the Wheeler-Feynman electromagnetic action-at-a-distance theory onto 
superspace was considered in Ref. \cite{TZh}. The classical vector and spinor fields belonging to 
the Maxwell supermultiplet were built from  the worldline coordinates of the charged and neutral particles in superspace. 
We believe that extension of these results to superstrings and super p-branes will make it possible 
to include the fermionic DOF in the processes of coalescence and fragmentation of super p-branes.

\vskip 10 pt

\noindent{\bf Acknowledgments}
\vskip 10 pt

The author would like to express his thanks 
to NORDITA and Physics Department of Stockholm University  
 for kind hospitality and support, to E. Bergshoeff, G. Huicken,
  H. Johansson, T. Koivisto, O. Lechtenfeld, S. Nicolis, A. Rosly, 
 H. von Zur-Muhlen and Yu. Zinoviev for interesting discussions.


\begin{thebibliography}{99}

\bibitem{RL}
F. Lund and T. Regge, 
Phys. Rev. D 14, 1524 (1976). 

\bibitem{BN}
B. M. Barbashov and V. V. Nesterenko, Introduction to the Relativistic 
String Theory (World Scientific, 1990).

\bibitem{Dirb}
J. Hoppe and H. Nicolai, 
Phys. Lett. B 196, 451 (1987).

\bibitem{Eisn}
L. F. Eisenhart,
Riemannian Geometry (Princeton Univ. Press, 1968).  

\bibitem{Car}
E. Cartan, Riemannian Geometry in an Orthogonal Frame 
(World Scientific, 2001). 

\bibitem{JS}
J. Simons,
Ann. Math. 88,  62 (1968);
R. Shoen, L. Simon and S. T. Yau,  
Acta. Math. 134, 275 (1975).

\bibitem{Vol1}
D. V. Volkov,  Phenomenological lagrangian of interaction for goldstone particles.
Kiev preprint ITF-69-75 (1969);  
Phys. of Elem. Part. At. Nucl. 4, 3 (1973).  
 
\bibitem{Wei}
S. Weinberg, 
Phys. Rev. Lett. 18,  188 (1967).   

\bibitem{Schw}
J. Schwinger,  
Phys. Lett. B 24, 473 (1967).     

\bibitem{CWZ}
S. Coleman, J. Wess and B. Zumino,
 Phys. Rev. 177, 2239 (1969).   

\bibitem{CCWZ}
C. Callan,  S. Coleman, J. Wess and B. Zumino, 
 Phys. Rev. 177, 2247 (1969).  

\bibitem{FST}
M. A. Semenov-Tyan-Shansky and L. D. Faddeev, 
Vestnik St. Petersburg Univ. 13, 81 (1977).  

\bibitem{Zgau2}
A. A. Zheltukhin, 
 Phys. Lett. B 116,  147 (1982); 
Theor. Math. Phys. 56, 785 (1983).   

\bibitem{Z_rmp}
A. A. Zheltukhin, 
 Rev. Math. Phys. 29,  1750009 (2017).  

\bibitem{Zgau_p}
A. A. Zheltukhin,
Phys. Part. Nucl. Lett. 14,  312 (2017);  
 Eur. Phys. J. C 74, 3048 (2014). 

\bibitem{GKW}
J. Gomis, K. Kamimura and P. West,
Class. Quant. Grav. 23, 7369 (2006).

\bibitem{GKP}
J. Gomis, K. Kamimura and J. M. Pons,
Nucl. Phys. B 871, 420 (2013). 

\bibitem{BZ_hedr}
I. A. Bandos and A. A. Zheltukhin, 
JETP Lett. 54,  421 (1991); 
 A. A. Zheltukhin,
 On the spinor structure of null strings and null membranes, 
preprint KhFTI-89-49, Kharkov KIPT (1989); https://lib-extopc.kek.jp/preprints/PDF/1989/8912/8912538.pdf

\bibitem{Stel}
 K. S. Stelle, 
 Int. J. Mod. Phys. A 32,  1741012 (2017); 
  Gen. Rel. Grav. 9, 353 (1978).

\bibitem{Strb}
  A. A. Starobinsky,
 Phys. Lett. B 91, 99 (1980). 

\bibitem{Frad}
E. S. Fradkin and A. A. Tseytlin,
 Phys. Lett. B 104, 377 (1981).

\bibitem{Zwi}
 B. Zwiebach, 
  Phys. Lett. B 156, 315 (1985).

\bibitem{LPPS}
H. L\"{u}, A. Perkins, C. N. Pope and K. S. Stelle,
Phys. Rev. D 92, 124019 (2015).  

\bibitem{TrZ} 
M. Trzetrzelewski and A. A. Zheltukhin,
  Phys. Lett. B 679, 523 (2009).   

\bibitem{Zcol} 
 A. A. Zheltukhin,	
Nucl. Phys. B 858, 142 (2012); 
 Nucl. Phys. B 867, 763 (2013).

\bibitem{Dir}
 P. A. M. Dirac,
  Proc. R. Soc. Lond. A 333, 403 (1973).

\bibitem{DFF}
V. De Alfaro, S. Fubini and G. Furlan,
Nuovo Cimento A 50, 523 (1979).

\bibitem{Smol}
L. Smolin,
Nucl. Phys. B 160, 253 (1979).  

\bibitem{Zee} 
A. Zee,
Phys. Rev. D 23, 858 (1980).  

\bibitem{Perv} 
V. Pervushin, A. Pavlov, 
Principles of Quantum Universe, LAP LAMBERT 
Academic Publishing, Saarbrucken, 2013.

\bibitem{GM}
F. Gliozzi, M. Meineri,
JHEP 1208, 056 (2012).

\bibitem{Zbc}
A. A. Zheltukhin, 
Sov. J. Nucl. Phys. 34, 311 (1981).
  
\bibitem{Shil}
A. Schild,
Phys. Rev. D 16, 1722 (1977).  

\bibitem{KL}
A. Karlhede and U. Lindstrom,
Class. Quant. Grav. 3, L73 (1986). 

\bibitem{Liz}
 F. Lizzi, G. Sparano, A. Srivastava and B. Rai, 
 Phys. Lett. B 182, 326 (1986).
 
\bibitem{GroM}
D. J. Gross and P. F. Mende, 
Phys. Lett. B 197, 129 (1987).

\bibitem{Znul}
A. A. Zheltukhin,
JETP Lett. 46, 262(1987); 
The Hamiltonian formalism for null strings, null membranes,
 null superstrings and null supermembranes,
 Preprint KFTI 87-46 (Kharkov, 1987).

\bibitem{BZnul}
I. A. Bandos and A. A. Zheltukhin,
Fortschr. Phys. 4, 619 (1993); 
 Phys. Part. Nucl. 25, 453 (1994). 

\bibitem{JoNo}
H. Johansson and J. Nohle, 
 Conformal Gravity from Gauge Theory,
 arXiv:1707.02965[hep-th].

\bibitem{RS}
 V. A. Rubakov and M. E. Shaposhnikov,
 Phys. Lett. B 125, 136 (1983). 

\bibitem{KR}
M. Kalb and P. Ramond,
Phys. Rev. D 9, 2273 (1974).  

\bibitem{YIIMYY}
M. Yamanobe, S. Ishikawa, Y. Iwama,
T. Miyazaki, K. Yamamoto and R. Yoshida,
 Prog. Theor. Phys. 96, 227 (1996). 

\bibitem{Z_nul}
A. A. Zheltukhin,
Sov. J. Nucl. Phys. 51,  950 (1990).
K. Ilienko and A. A. Zheltukhin,
 Class.Quant.Grav. 16, 383 (1999).  

\bibitem{VS}
D.V. Volkov and V.A. Soroka, 	
 JETP Lett. 18, 312 (1973);   
Theor. Mat. Phys. 20, 829 (1974).  

\bibitem{AFR}
R. d'Auria, P. Fre and T. Regge,
Group manifold approach to gravity and supergravity theories,
preprint ICTP, IC/81/54, MIRAMARE-TRIESTE, May 1981.  

\bibitem{Vol_Gap}
D. V. Volkov,
Generalized action principle for superstrings and supermembranes, 
 Proc. of the "SUSY-95" International Conference 
(Gif-sur-Yvette, Ed. Frontieres, 1996), 
arXiv: hep-th/9512103.

\bibitem{TZh}
V. V. Tugai and A. A. Zheltukhin,
Phys. Rev. D 51, 3997 (1995); 
ibid. D 54, 4160 (1996).  





\end{thebibliography}
\end{document}